%% file: main.tex
\documentclass[twocolumn,aps,prl,superscriptaddress,reprint,longbibliography]{revtex4-1}

\usepackage[latin9]{inputenc}
\usepackage{amsmath,tikz,graphicx}
\usepackage{esint}
\usepackage{wasysym}
\usepackage[colorinlistoftodos]{todonotes}
\makeatletter

\usepackage{H1}

\renewcommand{\ket}[1]{|{{#1}}\rangle}

\makeatother

\begin{document}

\title{Sub-ballistic growth of R\'enyi entropies due to diffusion}

\author{Tibor Rakovszky}

\affiliation{Department of Physics, T42, Technische Universit{\"a}t M{\"u}nchen, James-Franck-Stra{\ss}e 1, D-85748 Garching, Germany}

\author{Frank Pollmann}

\affiliation{Department of Physics, T42, Technische Universit{\"a}t M{\"u}nchen, James-Franck-Stra{\ss}e 1, D-85748 Garching, Germany}

\author{C.W.~von~Keyserlingk}

\affiliation{University of Birmingham, School of Physics \& Astronomy, B15 2TT,
UK}

\begin{abstract}

We investigate the dynamics of quantum entanglement after a global quench and uncover a qualitative difference between the behavior of the von Neumann entropy and higher R\'enyi entropies. We argue that the latter generically grow \emph{sub-ballistically}, as $\propto\sqrt{t}$, in systems with diffusive transport. We provide strong evidence for this in both a U$(1)$ symmetric random circuit model and in a paradigmatic non-integrable spin chain, where energy is the sole conserved quantity. We interpret our results as a consequence of local quantum fluctuations in conserved densities, whose behavior is controlled by diffusion, and use the random circuit model to derive an effective description. We also discuss the late-time behavior of the second R\'enyi entropy and show that it exhibits hydrodynamic tails with \emph{three distinct power laws} occurring for different classes of initial states.

\end{abstract}

\maketitle

\paragraph{Introduction.---} The far-from-equilibrium dynamics of closed quantum many-body systems has been at the center of much recent attention, both theoretically and experimentally~\cite{Rigol2008,CalabreseCardy06,ETHreviewRigol16,GogolinReview,Kaufman794}. In systems where the Eigenstate Thermalization Hypothesis (ETH)~\cite{Deutsch91,Srednicki94,ETHreviewRigol16} holds, the density matrix of a finite subsystem, $\rho_A$, relaxes to a Gibbs state with an extensive entropy that stems from the entanglement with the rest of the system, making the dynamics of entanglement integral to the understanding of equilibration. This question has recently become amenable to experimental probes in systems of cold atoms, through the measurement of \emph{R\'enyi entropies}, $S_\alpha \equiv \frac{1}{1-\alpha}\log{\text{tr}(\rho_A^\alpha)}$. The theoretically most relevant of these is the \emph{von Neumann entropy}, $S_{\alpha\to 1} \equiv - \text{tr}(\rho_A \log{\rho_A})$. Experimentally, however, for large subsystems only entropies with integer $\alpha \geq 2$ are currently accessible~\cite{DemlerEntanglement,ZollerEntanglement,Islam15,Kaufman794,Elben18}. It is therefore important to understand how their behavior might differ from that of $S_1$.

In generic clean systems, the von Neumann entropy is expected to grow linearly in time for approximately homogeneous initial states (`global quenches'). This is understood for integrable systems from a quasi-particle description~\cite{CalabreseCardy05,CalabreseCardy07,Alba2017_1,Alba2017_2}, but it also holds for thermalizing models~\cite{HyungwonHuse}, where it has recently been described using a `minimal cut' picture~\cite{Nahum16,JonayNahum,Mezei18}. A generic linear growth of $S_2$ has also been proposed in Refs. \onlinecite{Abanin17,Mezei16}, based on the ballistic spreading of operators -- this is consistent with existing results both in integrable systems~\cite{Fagotti2008,Alba2017_3,Mestyan_2018} and in models with no conservation laws~\cite{RvK17,Nahum17,ZhouNahum,Bertini2018}. Here we argue that this picture changes drastically in systems exhibiting diffusive transport of some conserved quantity (spin, charge, energy, etc.)~\cite{BLOEMBERGEN1949,DEGENNES1958,KADANOFF1963}: we find that $S_{\alpha > 1}$ grows \emph{diffusively}, as $\sqrt{t}$. This arises because entropies with $\alpha > 1$ are sensitive to the presence of a few anomalously large eigenvalues of the reduced density matrix, while $S_1$ is dominated by the many exponentially small eigenvalues. The possibility of such qualitative differences was discussed for global cat states in Ref. \onlinecite{JonayNahum}; here we propose that it arises much more generally, without the need to fine tune the initial state.
\begin{figure}[h!]
	\includegraphics[width=1.\columnwidth]{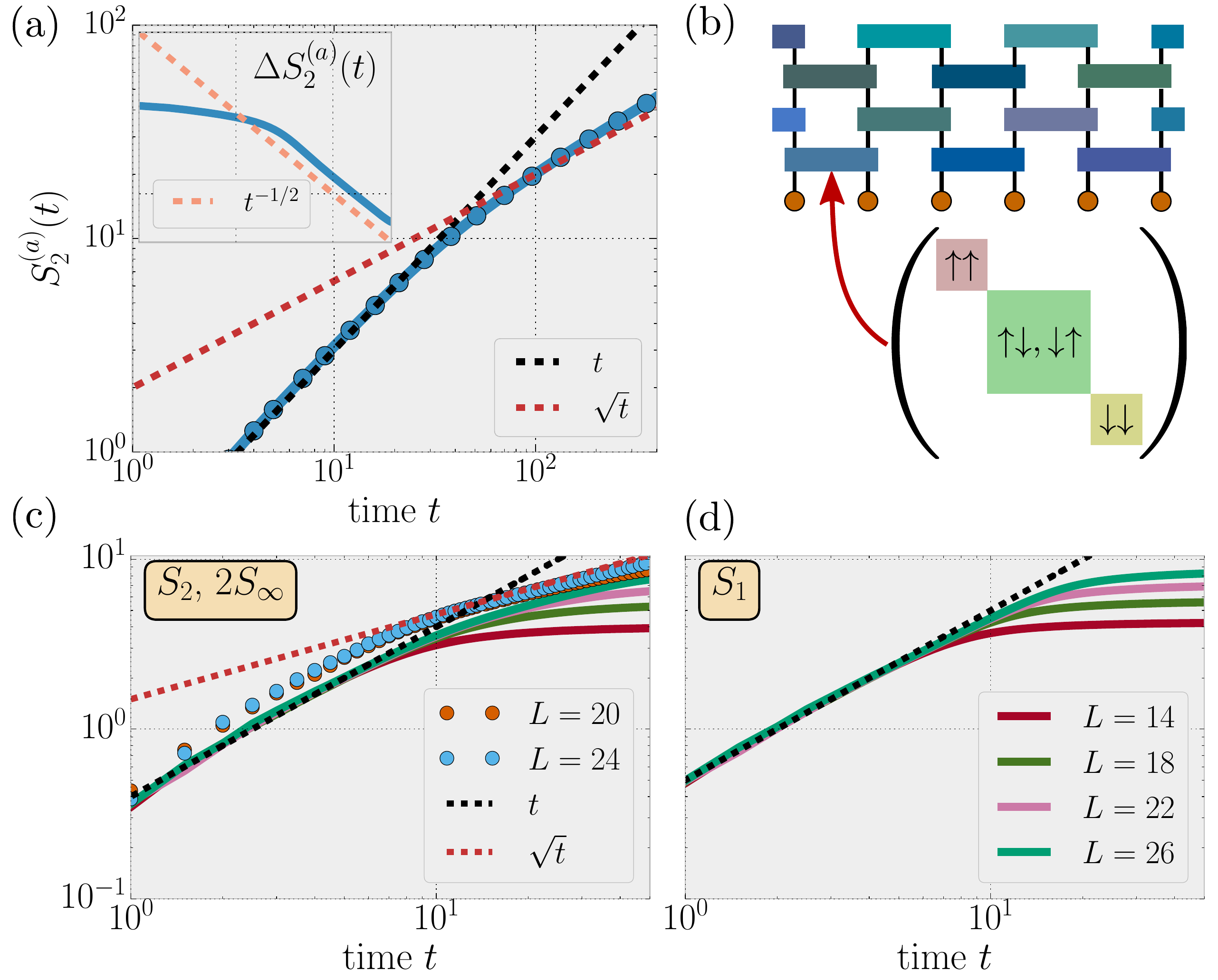}
	\caption{(a) Growth of (annealed average) second R\'enyi entropy in a spin $1/2$, U$(1)$-symmetric random circuit, averaged over all product states. At long times the growth is diffusive ($\propto\sqrt{t}$). Inset: the discrete time derivative $\Delta S_2^{(a)}(t) \equiv S_2^{(a)}(t+1)-S_2^{(a)}(t)$ decays as $t^{-1/2}$. (b) Geometry of the random circuit and block structure of the gates. (c): R\'enyi entropies of the tilted field Ising model~\eqref{eq:KimHuseDef}, $S_2$ (solid lines) and $2S_\infty$ (dots) show a similar cross-over to sub-ballistic growth, while (d) the von Neumann entropy grows mostly linearly.}  
	\label{fig:diffusive_growth}
\end{figure}
\paragraph{Numerical results.---}  We consider a local random unitary circuit with a conserved U$(1)$ charge as a minimal model of local quantum dynamics with diffusive transport~\cite{OTOCDiff1,OTOCDiff2}. We take a spin $1/2$ chain and evolve it with 2-site unitary gates that are block-diagonal in the total $z$-spin (Fig.~\ref{fig:diffusive_growth}b). Each unitary consists of three independent Haar random blocks, corresponding to the states $\{\uparrow\uparrow\}$, $\{\uparrow\downarrow,\downarrow\uparrow\}$, $\{\downarrow\downarrow\}$. In every time step we apply such two-site gates first on all even, then on all odd bonds of the chain, and the different gates in the circuit are all independently chosen. Clearly, this circuit conserves the total Pauli $z$ component, $\sum_x Z_x$.

We examine the circuit-averaged purity, $\overline{\mathcal{P}}$, where the purity is $\mathcal{P} \equiv e^{-S_2} = \text{tr}(\rho_A^2)$. This defines the \emph{annealed average} R\'enyi entropy, $S_2^{(a)}\equiv -\log{\overline{\mathcal{P}}}$, which lower bounds the average, $\overline{S_2} \geq S_2^{(a)}$. $\overline{\mathcal{P}}$ is represented as a classical partition function, using the mapping derived in Ref. \onlinecite{OTOCDiff2}, which we evaluate using standard tensor network methods~\cite{MurgReview,VidalTEBD}, making sure that the results are converged in both system size and bond dimension. Moreover, we average over all initial product states exactly. As shown in Fig.~\ref{fig:diffusive_growth}a, we find that at long times, the entanglement grows as $S_2^{(a)} \propto \sqrt{t}$. Note that the same quantity would grow linearly if we removed the conservation law~\cite{RvK17,Nahum17}, thus we attribute its slow growth to diffusive transport. 

Next, we consider the spin $1/2$ Hamiltonian 
\begin{equation}\label{eq:KimHuseDef}
H = J\sum_{r=1}^{L-1}  Z_r  Z_{r+1} + \sum_{r=1}^{L} (h_z  Z_r + h_x  X_r) - J ( Z_1 +  Z_{L}),
\end{equation}
known as the \emph{tilted field Ising model}. The last term is included to decrease boundary effects. We choose $J = 1$, $h_x = (5+\sqrt{5})/8$, and $h_z = (1+\sqrt{5})/4$, as the same model was previously shown to have diffusive energy transport and linear von Neumann entropy growth~\cite{HyungwonHuse}. Fig.~\ref{fig:diffusive_growth}c,d show the growth of different entropies, averaged over $N=50$ ($N=20$) random product states for system sizes $L=12-24$ ($L=26$). Here we average the entropies, not their exponentials, unlike the random circuit case. We observe a mostly linear growth of $S_1$, as in Ref. \onlinecite{HyungwonHuse}. $S_2$, however, has a cross-over to sub-linear growth at long times. Although the times we can reach are limited by finite system size, the long-time behavior is consistent with $S_2\propto\sqrt{t}$. The results become clearer when considering the \emph{min-entropy}, $S_{\alpha\to\infty}$, which provides an upper bound on $S_{\alpha > 1} \leq \frac{\alpha}{\alpha-1} S_\infty$. We find that $S_\infty$ is less sensitive to finite size effects, and exhibits a more pronounced cross-over towards $\sqrt{t}$ growth (dots in Fig.~\ref{fig:diffusive_growth}c). Similar results hold also for particular initial states, without averaging~\cite{suppmat}. The behavior of the random circuit and Hamiltonian models leads us to conjecture that diffusive growth of $S_{\alpha > 1}$ is a generic consequence of diffusive hydrodynamic transport. In the following we provide further justification of this conjecture.

\paragraph{Heuristic argument.---} We interpret our results in terms of the following non-rigorous argument. Let us focus on a $Z$-conserving discrete local time evolution, $U(t) = \prod_{\tau < t} U(\tau,\tau+1)$, on an infinite chain, and consider the bipartite entanglement at a cut between sites $x$ and $x+1$. We write the time evolved state as a `sum over histories', $\ket{\psi(t)} = \sum_{\{\sigma(\tau)\}} A(\{\sigma(\tau)\}) \ket{\sigma}$, where $A(\{\sigma(\tau)\})$ is the probability amplitude of a world history $\{\sigma(\tau)\}|_{0\leq \tau \leq t}$ in the $Z$ basis. We split this sum into two parts: i) histories for which the sites $x,x+1$ have both spins up at all times $\tau > t_\text{loc}$ for some local equilibration time $t_\text{loc}\sim\mathcal{O}(1)$, and ii) all remaining paths. Let $\ket{\phi_{0}(t)}$ and $\ket{\phi_{1}(t)}$ denote the normalized states corresponding to i) and ii) respectively. Then $\ket{\psi(t)}=c_{0}\ket{\phi_{0}(t)}+c_{1}\ket{\phi_{1}(t)}$. 
By construction, $\ket{\phi_0}$ has an $\mathcal{O}(1)$ Schmidt rank for a bi-partition across the bond $x,x+1$, accumulated before $t_\text{loc}$. Denoting this Schmidt rank by $\chi$, one can then use the \emph{Eckart-Young theorem}~\cite{EckartYoung,Huang2019} to lower bound the largest Schmidt value of $\ket{\psi}$ as
\begin{equation}\label{eq:Eckart}
\chi (\Lambda_\text{max}^\psi)^2 \geq |\braket{\phi_0|\psi}|^2 = |c_0 + c_1 \braket{\phi_0|\phi_1}|^2.
\end{equation}
We will now argue that if transport is diffusive, the RHS is expected to decay slower than exponentially with time. 

We first need to estimate the probability that the sites $x,x+1$ remain in the state $\uparrow\uparrow$ at all times $t>t_\text{loc}$. The simplest approximation is to treat every $\downarrow$ in the system as an independently diffusing particle. In this case, the probability that all particles that are to the left of $x$ at $t_\text{loc}$ remain on the same side is a product of the probabilities for particles starting at different positions. The relevant contribution comes from particles that are initially within some region of size $\mathcal{O}(\sqrt{Dt})$ near the entanglement cut, where $D$ is the diffusion constant. Therefore, we expect the probability to decay at long times as $|c_0|^2 \propto e^{-\gamma\sqrt{Dt}}$ for some constant $\gamma$~\cite{HardCoreNote}. 

To bound the overlap $\braket{\phi_0|\phi_1}$, we can apply the Eckart-Young theorem again, this time for $\ket{\phi_1}$, which gives $|\braket{\phi_0|\phi_1}|^2 \leq \chi (\Lambda_\text{max}^{\phi_1})^2$. Consequently, if $\ket{\phi_1}$, which corresponds to typical histories, has R\'enyi entropies $S_{\alpha > 1}$ that grow faster than $\sqrt{t}$, then the second term on the RHS of Eq.~\eqref{eq:Eckart} will be negligible at long times compared to $c_0$, resulting in $\chi (\Lambda_\text{max}^\psi)^2 \gtrsim e^{-\gamma \sqrt{Dt}}$. If $\Lambda_\text{max}^{\phi_1} \sim e^{-\sqrt{t}}$, then there is in principle a possibility of cancellation between the two terms, such that the RHS of Eq.~\eqref{eq:Eckart} decays faster then $\sim e^{-\sqrt{t}}$; however this would be highly fine-tuned and we see no sign of such cancellation when computing the RHS in the random circuit model.

This argument implies that at long times, there should be a growing distance between the largest Schmidt value, $\Lambda_\text{max} \equiv e^{-S_\infty/2} \sim e^{-\gamma\sqrt{Dt}}$, and typical ones which we still expect to be exponentially small, $\sim e^{-vt}$, in accordance with the linear growth of $S_1$~\cite{HyungwonHuse}. As mentioned previously, the former upper bounds R\'enyi entropies $S_{\alpha > 1} \leq \frac{\alpha}{\alpha-1} S_\infty$. This shows that at long times, $t \gg v^2/D$, all $\alpha > 1$ entropies are controlled by the largest Schmidt value, making their growth diffusive, provided that all degrees of freedom couple to some conserved quantity. The time for this sub-ballistic growth to set in depends also on the R\'enyi index, diverging in the limit $\alpha\to 1$. The von Neumann entropy itself is unconstrained by $S_\infty$, dominated instead by the many exponentially small Schmidt values, leading to its linear growth.

While the above argument is presented in the language of spin conservation, we expect it to generalize to energy conserving systems in the form of rare events where the time evolved state locally resembles the ground state~\cite{FrustFreeNote}. This is in agreement with our numerical results in Fig.~\ref{fig:diffusive_growth}c. 

\paragraph{Effective model---} To get a further analytical handle on this problem, we return to our random circuit model and modify it along the lines of Ref. \onlinecite{OTOCDiff1}, adding an extra, non-conserved $q$-state system to each site. This makes the size of each Haar random unitary larger by a factor of $q^2$, allowing us to derive an effective model that governs the evolution of $S_2^{(a)}$ in the large $q$ limit. As we shall see, the result decomposes into the sum of two contributions: a $\propto t$ part from the non-conserved degrees of freedom, and a $\propto\sqrt{t}$ part associated to the conserved spins.

\begin{figure}
	\includegraphics[width=1.\columnwidth]{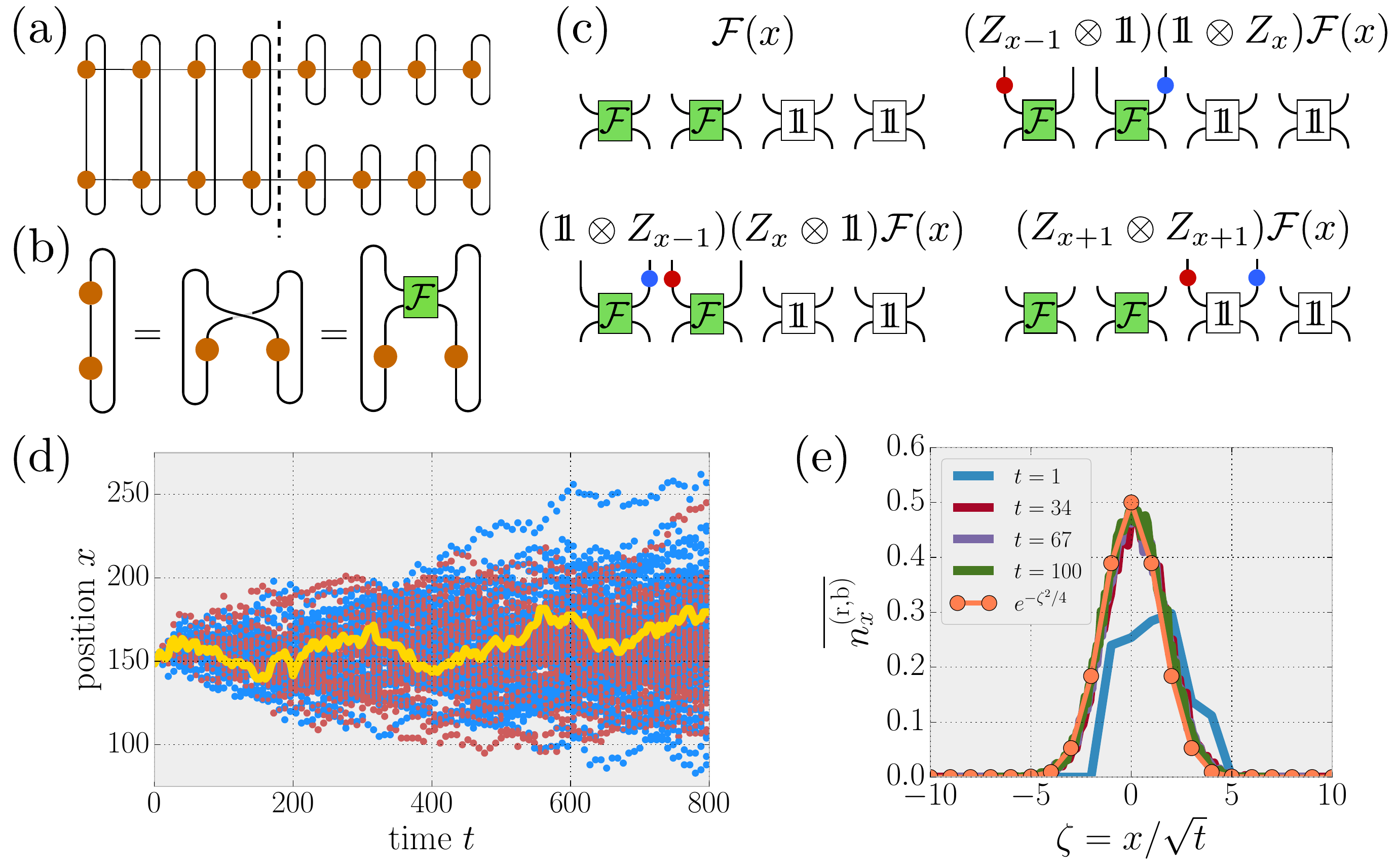}
	\caption{Effective model at $q\to\infty$. (a) The purity $\mathcal{P}$, written in terms of the state $\rho$ as a matrix product operator. (b) This can be rewritten by introducing the swap operator $\mathcal{F}$. (c) Half-chain `swap-string' $\mathcal{F}(x)$, along with a few of the terms on the RHS of Eq.~\eqref{eq:large_q_eom}, with red and blue particles representing local $ Z_x$ operators. (d) These particles obey a random walk with hard core interactions, spreading out diffusively, which (e) leads to a Gaussian density profile.}  
	\label{fig:redblue}
\end{figure}

To begin, note that the purity can be written as the expectation value of an operator on two identical copies of the original system~\cite{HastingsSwap,DemlerEntanglement,Hayden2016,Nahum17}. Imagine two copies of a single site, with Hilbert space $\mathcal{H} \otimes \mathcal{H}$, and define the one-site \emph{swap operator} $\mathcal{F}$, such that $\mathcal{F}(\ket{i} \otimes \ket{j}) \equiv \ket{j} \otimes \ket{i}$, where $\{\ket{i}\}$ is a basis in $\mathcal{H}$. Then the half chain purity is $\mathcal{P}(x) = \text{tr} (\mathcal{F}(x) [\rho \otimes \rho]) \equiv \braket{\mathcal{F}(x)}$, where $\mathcal{F}(x) \equiv \prod_{\leq x} \mathcal{F}$ is a string of swap operators, acting on one half of the entanglement cut (Fig.~\ref{fig:redblue}a,b). Instead of evolving the state, we can therefore evolve the operator $\mathcal{F}(x)$ in time. Averaging over a random gate on sites $x,x+1$, to leading order in $1/q$ it evolves as~\cite{suppmat}
\begin{equation}\label{eq:large_q_eom}
\mathcal{F}(x) \to (2q)^{-1} \sum_{y=x\pm 1} \mathcal{F}(y)+\tilde{\mathcal{F}}_x(y),
\end{equation}
where we have introduced new `two-copy' operators
\begin{equation}\label{eq:cut_dressed}
\tilde{\mathcal{F}}_{x}(y) \equiv (Z_{x,x+1} \otimes Z_{x,x+1}) \mathcal{F}(y),
\end{equation}
with $Z_{x,x+1} \equiv (Z_{x} +  Z_{x+1})/2$, and the tensor product referring to the two copies of the system. These are similar to $\mathcal{F}(x)$, but multiplied by the $Z$ operators that measure the conserved spin on sites near the entanglement cut.

To see how the entanglement evolves, one also needs equations of motion for $\tilde{\mathcal{F}}_{x}(y)$. This can be done analogously, by averaging over 2-site gates, resulting in the following effective model. The operators in  Eq.~\eqref{eq:cut_dressed}, and their subsequent circuit-averaged evolution, may be expressed as a sum of dressed swap operators of the form $\prod_y (Z_y^{n_y^{(r)}} \otimes Z_y^{n_y^{(b)}}) \mathcal{F}(x')$, where $\{n^{(r,b)}_y = 0,1\}$. We refer to $\{n^{(r,b)}_y\}$ as configurations of 'red' and 'blue' particles, while $x'$ (the endpoint of the swap string) is called the 'cut position'. Apart from an overall suppression factor of $2/q$ in each step, the circuit-averaged dynamics gives a Markov process on configurations defined by these variables. This effective Markov dynamics has the following properties: away from the cut, the particles independently obey diffusion with hard-core interactions, conserving the number of each species. The cut itself also diffuses, moving one site either to the left or the right, while emitting and absorbing an even number of particles at each step (Fig.~\ref{fig:redblue}d). One can show that the probability of emission vs. absorption decreases with the number of particles on the two sites directly at the cut, changing sign at half filling~\cite{suppmat}. 

The swap-string and both types of particles evolve as unbiased random walks, therefore by time $t$ we expect them to occupy a region of width $l(t)\propto\sqrt{t}$. Monte Carlo simulations of the stochastic dynamics show (Fig.~\ref{fig:redblue}e) that the particle densities are Gaussian around $x$. We therefore take a mean field approximation and write the probability of a string ending at $x$ and a distribution of particles $\{n^{(r,b)}_y = 0,1\}$ as
\begin{equation}
p(x,n^{(r)},n^{(b)})\propto e^{-\frac{x^{2}}{2l(t)^2}} e^{-\frac{1}{2l(t)^2}\sum_{y}\left(n_{y}^{(r)}+n_{y}^{(b)}\right)y^{2}}
\end{equation}
if $\sum_{y}(n_{y}^{(r)}+n_{y}^{(b)})$ is even, and zero otherwise. With this ansatz, one can evaluate the half-chain purity at time $t$. For translation invariant product states, the result reads
\begin{equation}\label{eq:large_q_purity_final}
\mathcal{P}(t) \propto\left(\frac{2}{q}\right)^t \times \prod_{y}\left(1-\frac{1-\left|\braket{ Z_{y}}\right|}{1+e^{y^{2}/2l(t)^2}}\right)^{2}.
\end{equation}
This product has a relevant contributions only from a window of
$|y| \lesssim \sqrt{t}$, hence it decays as $e^{-\gamma\sqrt{t}}.$
Note that $\gamma$ is larger when $|\braket{ Z_{y}}|$ is smaller. By expanding each term in $e^{-y^2/2l(t)^2}$, we can approximate the product as $\approx e^{-{\sqrt{2\pi}(1-|\braket{ Z}|) l(t)}}$, which is in good agreement with Monte Carlo results, at least away from $\braket{ Z} \approx 0$. Note that Eq.~\eqref{eq:large_q_purity_final} looks very similar to the probability of rare events from our heuristic argument in the simplest approximation of independently diffusing particles.

Using Eq.~\eqref{eq:large_q_purity_final}, at large $q$ we get $S_2^{(a)} = \log(q/2)\,t + a\sqrt{t}$, with $a \sim \mathcal{O}(1)$. Here, $\log(q/2)$ is exactly the large $q$ limit of $v_\text{E}(q) = \log{\frac{q+1/q}{2}}$, the entanglement velocity of a non-symmetric random circuit with $q$ states per site~\cite{RvK17,Nahum17}. Moreover, the linear in $t$ term is independent of the initial state. This suggests the following interpretation: there is an entanglement $v_\text{E}(q)t$ coming entirely from non-conserved degrees of freedom, while the conserved spins are responsible for the $\propto\sqrt{t}$ term. This is supported by numerical results~\cite{suppmat}, which show that $S_2^{(a)}(t) - v_\text{E}(q)t$ has only weak $q$-dependence and grows as $\sqrt{t}$ for any $q$, including the original model with $q=1$, where $S_2^{(a)}$ is purely diffusive (Fig.~\ref{fig:diffusive_growth}a). This is true despite the ballistic spread of local operators~\cite{OTOCDiff1,OTOCDiff2}, showing that recent arguments~\cite{Abanin17,Mezei16,RvK17} for exponential decay of $\mathcal{P}$ fail in the present context, and subtle correlations between the spreading of different operators cannot be neglected. Our results also suggest that the `minimal cut' picture of entanglement growth~\cite{JonayNahum} does not accurately capture the behavior of $S_{\alpha > 1}$~\cite{MarkNote}.

\paragraph{Long-time tails.---} Diffusive modes also have a strong influence on the long-time behavior of finite subsystems, which we turn to next. The entanglement eventually saturates to an equilibrium value predicted by the appropriate Gibbs ensemble, provided ETH holds and the initial state clusters~\cite{Deutsch91,Srednicki94,Rigol2008,ETHreviewRigol16}. We now show that the approach of $S_2$ to this thermodynamic value is also affected by diffusion and shows long-time hydrodynamic tails. Interestingly, we find that the nature of these tails depends strongly on the initial conditions, leading to the appearance of three different power laws, $t^{-1/2}$, $t^{-1}$ and $t^{-3/2}$. In particular we uncover a difference between states at zero and finite chemical potential.

We take a spin $1/2$ chain and rewrite the reduced density matrix of a small subsystem of $l$ sites by inserting a complete basis of operators $\sigma^\mu$, given by  products of local Pauli operators acting on the subsystem~\cite{Abanin17}. This yields $S_2 = l \log{2} - \log\left(1 + \sum_\mu \braket{\sigma^\mu}^2\right)$, where the identity is excluded from the sum. Let $\braket{\delta \sigma^\mu} \equiv \braket{\sigma^\mu} -\braket{\sigma^\mu}_\text{eq}$ denote the deviation from equilibrium. Then at long times
\begin{equation}\label{eq:entanglement_saturation}
|S_2 - S_{2,\text{eq}}| \propto \sum_\mu \left( 2 \braket{\sigma^\mu}_\text{eq} \braket{\delta \sigma^\mu} + \braket{\delta \sigma^\mu}^2 \right).
\end{equation}
Thus the long-time tails that describe how expectation values equilibrate appear directly in the R\'enyi entropy.

One immediate consequence of Eq.~\eqref{eq:entanglement_saturation} is that the hydrodynamic tails can differ between states at half filling and away from half filling. At precisely half filling, the leading order term is $\braket{\delta \sigma^\mu}^2$, while away from half filling $\braket{\sigma^\mu}_\text{eq}\braket{\delta \sigma^\mu}$ is expected to dominate. Generically, hydrodynamic observables in $d$ dimension should decay as $t^{-d/2}$ ~\cite{BLOEMBERGEN1949,DEGENNES1958,KADANOFF1963,Rosch13}, with subleading corrections $\mathcal{O}(t^{-3d/4})$. Therefore, we generically expect a saturation as $\propto t^{-d}$ for states at half filling (infinite temperature) and $\propto t^{-d/2}$ otherwise. However, this expectation can change for certain initial states, where all hydrodynamic variables have $\braket{\delta\sigma^\mu} = 0$ initially. In this case one expects the leading diffusive tail to vanish and subleading corrections to take over. In particular, in the 1D random circuit model one can argue~\cite{suppmat} that the leading contribution for translation invariant product states should be of order $t^{-3/2}$. 

\begin{figure}[t!]
	\includegraphics[width=1.\columnwidth]{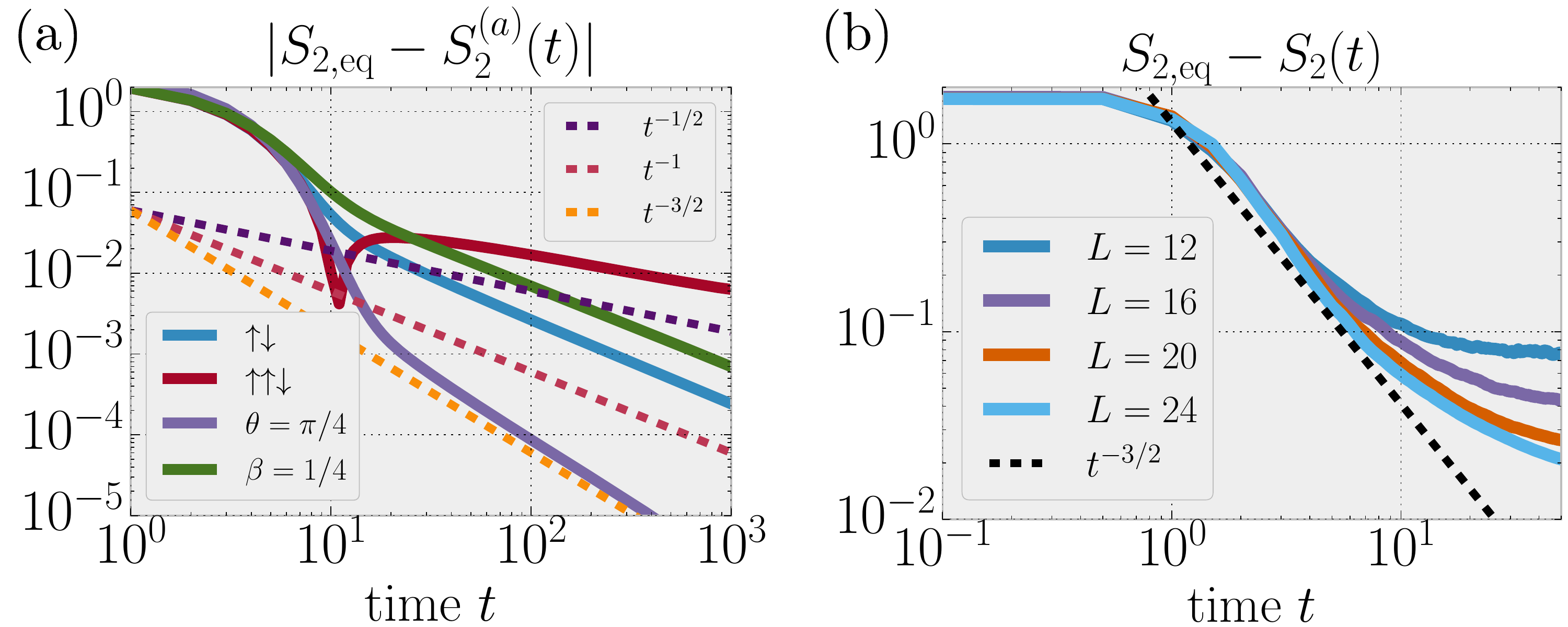}
	\caption{(a) Saturation of $S_2^{(a)}$ for a 4-site subsystem in the spin $1/2$ random circuit for different initial states. States away/at half filling generically saturate as $t^{-1/2},t^{-1}$ respectively. States with hydrodynamic variables in equilibrium at all times  saturate with subleading exponent $t^{-3/2}$. (b): The same $t^{-3/2}$ saturation is present for random product states in the tilted field Ising model~\eqref{eq:KimHuseDef} (3-site subsystem, averaged over $50$ initial states).} 
	\label{fig:entanglement_sat_RC}
\end{figure}

We observe these three distinct power laws in $S_2^{(a)}$ for the spin $1/2$ random circuit, as shown in Fig.~\ref{fig:entanglement_sat_RC}a. We find that N\'eel-like states ($\ket{\uparrow\uparrow\downarrow\uparrow\uparrow\downarrow\ldots}$) with less then half filling exhibit an overshooting effect, approaching their equilibrium value \emph{from above}, as $t^{-1/2}$. Finitely correlated states at half filling ($\ket{\beta} \propto e^{\beta \sum_{r=0}^{L-1}  Z_r  Z_{r+1}} \left(\ket{\uparrow} + \ket{\downarrow} \right)^{\otimes L}$) saturate as $t^{-1}$, and tilted ferromagnetic states ($\ket{\theta} \equiv e^{i \theta \sum_{r=1}^{L} Y_r} \ket{\uparrow}$) as $t^{-3/2}$. We also provide evidence of the $t^{-3/2}$ tail for random product states in the tilted field Ising Hamiltonian~\eqref{eq:KimHuseDef}, shown in Fig.~\ref{fig:entanglement_sat_RC}b.

\paragraph{Discussion.---} Our results reveal a previously overlooked qualitative difference between the von Neumann and $\alpha > 1$ R\'enyi entropies. We gave a heuristic argument, indicating that the latter are strongly influenced by local quantum fluctuations which can lead to diffusive growth for the entropy in systems with diffusive transport. We presented evidence for this in two cases: a random circuit model and a thermalizing Hamiltonian. Our results indicate that diffusion leads to a separation of scales, where the half-chain density matrix contains a few largest eigenvalues that decay slowly and become well-separated from the bulk of the spectrum made up by exponentially small eigenvalues: $S_2$ is dominated by the former, while $S_1$ by the latter, such that they provide insight into different aspects of thermalization. In particular, $S_1$ and $S_{\alpha>1}$ reveal distinct time scales of thermalization, respectively scaling as $\propto l$ and $\propto l^2$ with subsystem size $l$.

We expect our results to generalize to higher dimensions. In that case, the swap-string becomes a `membrane'~\cite{Nahum16,JonayNahum,Mezei18}. One can generate a 2D time evolution using a random circuit of 2-site gates~\cite{Nahum17}, in which case our Eq.~\eqref{eq:large_q_eom} remains valid, with the membrane emitting $ Z_x$ operators that diffuse on the 2D lattice. A generalization of our heuristic argument would also suggest a similarly sub-ballistic growth for the R\'enyi entropy. It would be interesting to see if effects like this could show up in holographic calculations, by extending the results of Ref. \onlinecite{Mezei17,Mezei18} to R\'enyi entropies, or by refining the upper bound derived for SYK chains in Ref. \onlinecite{Gu2017}

It is an open question whether diffusion also affects the early-time growth of $S_1$, e.g. the form of subleading corrections. While slow relaxation of $S_1$ due to diffusion has been observed numerically~\cite{KimHuse2015}, whether the rich variety of initial state-dependent power laws appear there also warrants further study. Another avenue for future investigation is in the field of disordered systems, where even the von Neumann entropy is expected to grow sub-ballistically, while transport becomes sub-diffusive~\cite{Znidaric2016,Luitz2016,Nahum18}, eventually leading to many-body localization at strong disorder. Comparison of von Neumann and R\'enyi entropies could give further insight into the dynamics in these different regimes. 

\acknowledgements{We thank Michael Knap for useful comments. We are especially grateful to Adam Nahum for an illuminating discussion concerning the effect of rare events on entanglement growth. FP acknowledges the support of the DFG Research Unit FOR 1807 through grants no. PO 1370/2- 1, TRR80, the Nanosystems Initiative Munich (NIM) by the German Excellence Initiative, and the European Research Council (ERC) under the European Unions Horizon 2020 research and innovation program (grant agreement no. 771537). CvK is supported by a Birmingham Fellowship.}

\textbf{Note Added.} After posting the manuscript, a follow-up work~\cite{Huang2019} appeared, which provided a proof of sub-ballistic growth of $S_{\alpha > 1}$ for charge-conserving unitary circuits, under slightly different assumptions, and for a specific set of initial states. This work made us aware of the Eckart-Young theorem, which allowed us to tighten our heuristic argument, replacing one of its underlying assumptions with a milder condition.

\bibliographystyle{apsrev4-1}
\input{main.bbl}

\onecolumngrid
\appendix

\begin{center}
{\large \textbf{Supplementary Material for ``Sub-ballistic growth of R\'enyi entropies due to diffusion''\\}}
\end{center}

\section{Circuit-averaged purity dynamics}

We first re-derive the mapping from the random circuit model to an effective `classical' description, originally described in Ref. \cite{OTOCDiff2}, generalizing it to the model described in Ref. \cite{OTOCDiff1} in the process. This mapping provides the basis for an efficient method for calculating the average purity of the time evolved state, which we make us of in our numerical calculations. Furthermore, we take the $q\to\infty$ limit to arrive at the rules of the effective model presented in the main text.

We are interested in evolving operators that act on two copies of the original Hilbert space, which are relevant for evaluating the average purity. Time evolving such two-copy operators involves four instances of the time evolution operator $U$, therefore we require the fourth moment of each random gate, which can be evaluated in a relatively straightforward way. The result of averaging over a two-site gate is then a four-leg tensor. Each leg of this tensor in principle corresponds to $\mathcal{H}_1^{\otimes 4}$, where $\mathcal{H}_1$ is the one-site Hilbert space; however, in practice only a smaller dimensional subspace of this is relevant after averaging. Computing the average purity can then be achieved by contracting all these tensors together, which defines a two-dimensional tensor network, analogous to the calculation of the partition function of a 2D classical spin system. The boundary conditions of this partition function at times $0$ and $t$ depend on the initial state $\rho(0)$ and on the choice of the subsystem $A$, respectively.

We are interested in evaluating this partition function taking the on-site Hilbert space $\mathbb{C}^2 \otimes \mathbb{C}^q $ described in Ref. \cite{OTOCDiff1}. In what follows $Z_x$ denotes the local value of the conserved spin, which is an operator acting only on the $\mathbb{C}^2$ component of the on-site Hilbert space. It will not be  necessary to construct an explicit operator basis for the $\mathbb{C}^q$ degrees of freedom. For arbitrary pairs of operators $u,v$, acting on a single site in the original Hilbert space $\mathbb{C}^2 \otimes \mathbb{C}^q $, we can assign the following operators acting on the doubled Hilbert space

\begin{align}\label{eq:Lambdadef}
    \Lambda_{u\mid v}^{-}\equiv u\otimes v & & \Lambda_{u\mid v}^{+}\equiv  \left(u\otimes v\right) \mathcal{F},
\end{align}
where $\mathcal{F}= \sum_{i j} e_{i j} \otimes e_{j i} $ is the on-site swap operator, where $i$ is any on-site orthonormal basis. The tensor multiplication implicit in the second part of the above expressions is $(u\otimes v) (u'\otimes v') =u u'\otimes v v' $. The notion of Frobenius inner product between operators generalizes naturally to this two-copy space and yields
\begin{align}\label{eq:innerproducts}
\braket{\Lambda_{u\mid v}^{\pm}\mid\Lambda_{u'\mid v'}^{\pm}} = \text{tr}\left(u^{\dagger}u'\right)\text{tr}\left(v^{\dagger}v'\right) & & 
\braket{\Lambda_{u\mid v}^{\mp}\mid\Lambda_{u'\mid v'}^{\pm}}= & \text{tr}\left(v^{\dagger}v'u^{\dagger}u'\right).
\end{align}
The `purity operator' for the half-chain containing sites $\leq x$ then reads $\mathcal{F}(x)\equiv\Lambda_{1\mid1}^{\leq x,+}\otimes\Lambda_{1\mid1}^{>x,-}$. 

Consider now the effect of a two-site U$(1)$-symmetric unitary gate, acting on sites $x,x+1$. The two-site Hilbert space decomposes into three subspaces, labeled by the total charge $Q=0,1,2$, corresponding to total spin-$z$ components $+2,0,-2$ respectively. Let $P_{Q}$ denote the projector onto the two-site Hilbert spaces with charge $Q=0,1,2$, which have dimension $d_Q = q^2,2q^2, q^2$ respectively. Then averaging over four moments of the random gate yields an effective two-site evolution operator,
\begin{equation}\label{eq:avg_gate}
\overline{U^* \otimes U \otimes U^* \otimes U} \equiv \mathfrak{T}_{x,x+1}=  \sum_{\sigma,\mu=\pm1}\sum_{e_{1}e_{2}}\frac{w_{\sigma\mu}(e_{1},e_{2})}{d_{e_{1}}d_{e_{2}}-\delta_{e_{1}e_{2}}}\mid\Lambda_{P_{e_{1}}\mid P_{e_{2}}}^{\sigma}\rangle\langle\Lambda_{P_{e_{1}}\mid P_{e_{2}}}^{\mu}\mid
\end{equation}
where $w_{+}=1$ and $w_{-}=-\delta_{e_{1}e_{2}}d_{e_{1}}^{-1}$. The terms appearing in this effective evolution are two-copy operators acting on the pair of sites $x,x+1$. In order to contract the tensor network one then needs to split them up to a sum of single-site two-copy operators, which in principle can be done in a many different ways, depending on the choice of local basis. Given such a local basis choice, the computation of the average purity reduces to contracting $2t$ layers of such two-site tensors, along with the boundary conditions defined by $\rho_0 \otimes \rho_0$ (where $\rho_0$ is the initial density matrix) and $\mathcal{F}(x)$. The contraction can be done in a variety of ways. In our numerical calculations we use a boundary MPS method~\cite{MurgReview}, wherein the boundary tensors defined by $\mathcal{F}(x)$ are written as a matrix product state (MPS) and evolved layer-by-layer as in the Time Evolving Block Decimation (TEBD) algorithm~\cite{VidalTEBD}. Calculting the purity then amounts to taking the overlap of this time evolved MPS with another one that represents $\rho_0 \otimes \rho_0$. A further adventage of this method is that one can average over all product states analytically and incorporate the result into the boundary conditions, which yields the curve shown e.g. in Fig.~\ref{fig:diffusive_growth}a.

\subsection{Large $q$ limit}

\begin{figure}[h!]
    \centering
	\includegraphics[width=1.\columnwidth]{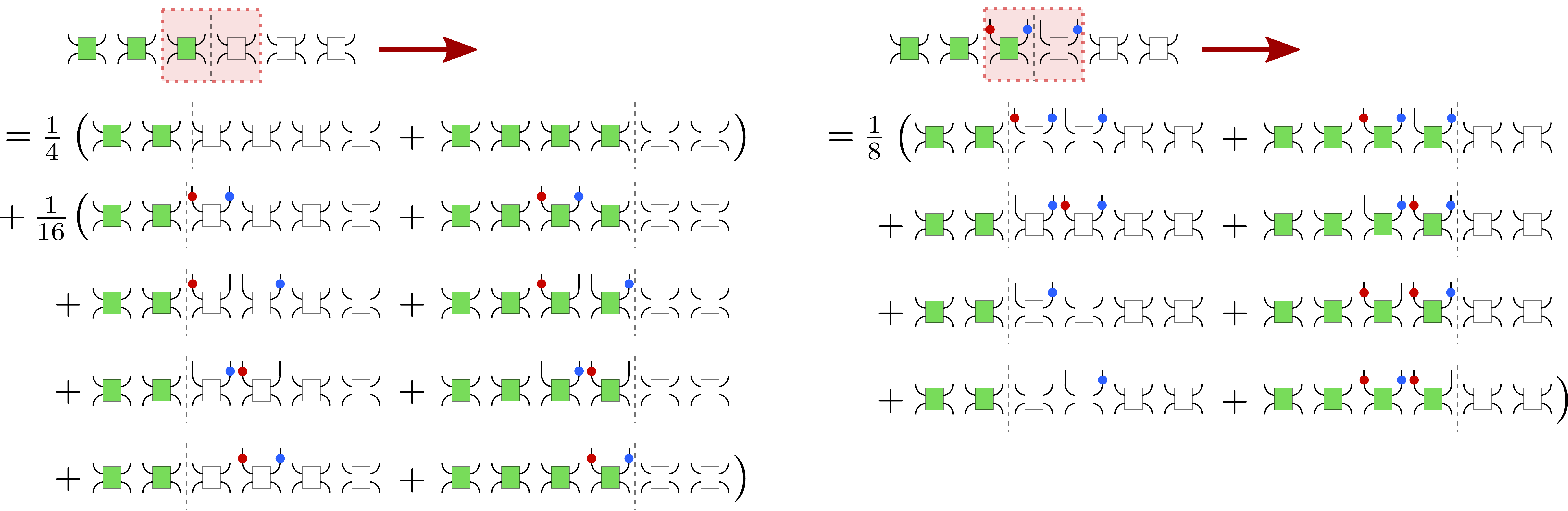}
	\caption{Illustration of some of the two-site update rules in the large $q$ effective model. Left: evolution of a swap-string, given by Eq.~\eqref{eq:entanglementeom}. The two-site gate (red dotted line) moves the string by one site, while producing spin-operators on both copies of the Hilbert space, denoted by red and blue particles respectively. Right: evolution of a more complicated configuration, with several particles already present near the endpoint of the string.} 
	\label{fig:Purity_EOM}
\end{figure}

While Eq.~\eqref{eq:avg_gate} yields an effective evolution that is numerically efficient, it is not analytically solvable, unlike the case of a random circuit without symmetries~\cite{Nahum18}. One can, however, simplify the equations by taking the $q\to\infty$ limit, which yields
\begin{equation}
\mathfrak{T}_{x,x+1}\left[\Lambda_{1\mid1}^{+,x}\otimes\Lambda_{1\mid1}^{-,x+1}\right]=\frac{2}{q}\times\frac{1}{2}\sum_{\sigma=\pm}\frac{1}{2}\left(\Lambda_{1\mid1}^{\sigma}+\Lambda_{Z_{x,x+1}\mid Z_{x,x+1}}^{\sigma}\right)+\mathcal{O}(q^{-3}).
\end{equation}
The terms $\Lambda_{Z_{x,x+1}\mid Z_{x,x+1}}^{\pm}$ are defined according to $Z_{x,x+1}= 2^{-1}(Z_x + Z_{x+1})$ and Eq.~\ref{eq:Lambdadef} and are identical to $\tilde{\mathcal{F}}_{x\mp 1}(x\pm 1)$ defined in the main text. Applied to the purity operator, the above equation then yields
\begin{equation}\label{eq:entanglementeom}
\mathfrak{T}_{r,r+1}\left[\mathcal{F}(x)\right]=\delta_{x\ne r}\mathcal{F}(x)+\delta_{x,r}\frac{2}{q}\times\frac{1}{4}\sum_{\sigma}\left(\mathcal{F}(x+\sigma)+\sum_{z,w=x,x+1}\frac{1}{4}\mathcal{F}_{z,w}(x+\sigma)\right)+\mathcal{O}(q^{-3}).
\end{equation}
Here we have defined 
\begin{equation}\label{eq:manifoldofsuperops}
\mathcal{F}_{\tau,\upsilon}(x)=\prod_{i}\Lambda_{Z_{i}^{\tau_{i}}\mid Z_{i}^{\upsilon_{i}}}^{(-1)^{\delta_{x <  i} }},    
\end{equation}
 $\tau_{i},\upsilon_{i}=0,1$ indicate the positions of the $z$ operators which we refer to as the positions of red and blue particles in the main text, as in Fig.~\ref{fig:redblue}c,d. $x$ is the position of the cut. We have abbreviated a special case of such an operator with notation $\mathcal{F}_{y,z}(x)$ where $x,y,z$ simply denote the position of the cut, and $y,z$ are the positions at which a single $\tau,\upsilon$ is nonzero. Eq.~\eqref{eq:entanglementeom} is exactly the result stated in Eq.~\eqref{eq:large_q_eom}. It is illustrated in terms of the red and blue particle picture in Fig.~\ref{fig:Purity_EOM}

Before moving on to derive equations of motion for these operators, let us try to gain some intuition about their physical meaning. When evaluated in a pure state they give, focusing on the simplest case,
\begin{equation}\label{eq:disting}
\braket{\tilde{\mathcal{F}}_{x-1}(x)} \equiv \langle \tilde{\mathcal{F}}_{x-1}(x)\mid \rho \otimes \rho\rangle = {\Big |}{\Big |} \text{tr}_{\leq x} \left( \hat P_0^{x-1,x} \rho - P_2^{x-1,x} \rho \right) {\Big |}{\Big |}^2,
\end{equation}
where $P_Q^{x-1,x}$ is a projector that projects onto eigenstates of the two-site charge operator $\hat Q_{x-1,x}$ with eigenvalue $Q$. This expression is a direct measure of how distinguishable the positive matrices $\hat P_{0}\rho \hat P_{0}$ and $\hat P_{2}\rho \hat P_{2}$ are through measurements purely on the subsystem consisting of sites to the right of $x$. It is therefore directly related to local quantum fluctuation of the charge density on these two sites. In particular they are sensitive to the sort of `rare configurations' discussed in our heuristic argument, where the neighborhood of the entanglement cut if completely empty/filled.

To find the equations of motion for the terms $\mathcal{F}_{\tau,\upsilon}(y)$, it is necessary and sufficient to determine the dynamics of various two site operators of form $\Lambda^{\pm,x}_{a\mid b} \otimes \Lambda^{\pm,x+1}_{a'\mid b'}$  and $\Lambda^{\pm,x}_{a\mid b} \otimes \Lambda^{\mp,x+1}_{a'\mid b'}$ where $a,b,a',b=1,Z'$.
To begin, the following operators are exactly invariant under $\mathfrak{T}_{x,x+1}$
\begin{align*}
\Lambda_{1\mid1}^{\pm,x}\otimes\Lambda_{1\mid1}^{\pm,x+1} & &
\Lambda_{Z_{x}\mid1}^{\pm,x}\otimes\Lambda_{Z_{x+1}\mid1}^{\pm,x+1} & &
\Lambda_{1\mid Z_{x}}^{\pm,x}\otimes\Lambda_{1\mid Z_{x+1}}^{\pm,x+1} & & \Lambda_{Z_{x}\mid Z_{x}}^{\pm,x}\otimes\Lambda_{Z_{x+1}\mid Z_{x+1}}^{\pm,x+1},
\end{align*}
These correspond to the statement that, far away from the cut, there are no dynamics for red (blue) particles if the pair of sites in question are empty or fully occupied with red (blue) particles. 

The following superoperators have very simple evolutions under $\mathfrak{T}_{x,x+1}$ which follow immediately from Haar averaging.
\begin{align*}
\Lambda_{Z_{x}\mid1}^{\pm,x}\otimes\Lambda_{1\mid1}^{\pm,x+1},\Lambda_{1\mid1}^{\pm,x}\otimes\Lambda_{Z_{x+1}\mid1}^{\pm,x+1} & \rightarrow\frac{1}{2}\left(\Lambda_{Z_{x}\mid1}^{\pm,x}\otimes\Lambda_{1\mid1}^{\pm,x+1}+\Lambda_{1\mid1}^{\pm,x}\otimes\Lambda_{Z_{x+1}\mid1}^{\pm,x+1}\right)\\
\Lambda_{1\mid Z_{x}}^{\pm,x}\otimes\Lambda_{1\mid1}^{\pm,x+1},\Lambda_{1\mid1}^{\pm,x}\otimes\Lambda_{1\mid Z_{x+1}}^{\pm,x+1} & \rightarrow\frac{1}{2}\left(\Lambda_{1\mid Z_{x}}^{\pm,x}\otimes\Lambda_{1\mid1}^{\pm,x+1}+\Lambda_{1\mid1}^{\pm,x}\otimes\Lambda_{1\mid Z_{x+1}}^{\pm,x+1}\right)\\
\Lambda_{Z_{x}\mid Z_{x}}^{\pm,x}\otimes\Lambda_{1\mid Z_{x+1}}^{\pm,x+1},\Lambda_{1\mid Z_{x}}^{\pm,x}\otimes\Lambda_{Z_{x+1}\mid Z_{x+1}}^{\pm,x+1} & \rightarrow\frac{1}{2}\left(\Lambda_{Z_{x}\mid Z_{x}}^{\pm,x}\otimes\Lambda_{1\mid Z_{x+1}}^{\pm,x+1}+\Lambda_{1\mid Z_{x}}^{\pm,x}\otimes\Lambda_{Z_{x+1}\mid Z_{x+1}}^{\pm,x+1}\right)\\
\Lambda_{Z_{x}\mid Z_{x}}^{\pm,x}\otimes\Lambda_{Z_{x+1}\mid1}^{\pm,x+1},\Lambda_{Z_{x}\mid1}^{\pm,x}\otimes\Lambda_{Z_{x+1}\mid Z_{x+1}}^{\pm,x+1} & \rightarrow\frac{1}{2}\left(\Lambda_{Z_{x}\mid Z_{x}}^{\pm,x}\otimes\Lambda_{Z_{x+1}\mid1}^{\pm,x+1}+\Lambda_{Z_{x}\mid1}^{\pm,x}\otimes\Lambda_{Z_{x+1}\mid Z_{x+1}}^{\pm,x+1}\right)
\end{align*}
We also meet more four more involved products
\begin{align*}
\Lambda_{Z_{x}\mid Z_{x}}^{\pm,x}\otimes\Lambda_{1\mid1}^{\pm,x+1},\Lambda_{1\mid1}^{\pm,x}\otimes\Lambda_{Z_{x+1}\mid Z_{x+1}}^{\pm,x+1} & \rightarrow\Lambda_{Z_{x,x+1}\mid Z_{x,x+1}}^{\pm}+\frac{1}{2q^{2}}\Lambda_{\frac{1-Z_{x}Z_{x+1}}{2}\mid\frac{1-Z_{x}Z_{x+1}}{2}}^{\mp}+\mathcal{O}_\text{rel}(q^{-4})\\
\Lambda_{Z_{x}\mid1}^{\pm,x}\otimes\Lambda_{1\mid Z_{x+1}}^{\pm,x+1},\Lambda_{1\mid Z_{x}}^{\pm,x}\otimes\Lambda_{Z_{x+1}\mid1}^{\pm,x+1} & \rightarrow\Lambda_{Z_{x,x+1}\mid Z_{x,x+1}}^{\pm}-\frac{1}{2q^{2}}\Lambda_{\frac{1-Z_{x}Z_{x+1}}{2}\mid\frac{1-Z_{x}Z_{x+1}}{2}}^{\mp}+\mathcal{O}(q^{-4}).
\end{align*}
The $O(q^{-2})$ terms on the second line are subleading, and introduce a new entanglement cut at the next time step, which will accrue additional suppressing $O(q^{-1})$ factors; we henceforth ignore these terms. The above 16 mappings reflect the statement that, far from the cut and at leading order, red and blue particles undergo single-file diffusion. 

We have listed all the possible products on one site of a cut. Finally we meet products involving sites near the entanglement cut, such as
\begin{align*}
\Lambda_{1\mid1}^{\pm,x}\otimes\Lambda_{1\mid1}^{\mp,x+1} & \rightarrow\frac{2}{q}\times\frac{1}{2}\sum_{\sigma=\pm}\frac{1}{2}\left({\Lambda_{1\mid1}^{\sigma}}+{\Lambda_{Z_{x,x+1}\mid Z_{x,x+1}}^{\sigma}}\right)+O(q^{-3})\\
\Lambda_{Z_{x}\mid1}^{\pm,x}\otimes\Lambda_{Z_{x+1}\mid1}^{\mp,x+1} & \rightarrow\frac{2}{q}\times\frac{1}{2}\sum_{\sigma=\pm}\frac{1}{2}\left({\Lambda_{Z_{x}Z_{x+1}\mid1}^{\sigma}+\Lambda_{Z_{x,x+1}\mid Z_{x,x+1}}^{\sigma}}\right)+O(q^{-3})\\
\Lambda_{1\mid Z_{x}}^{\pm,x}\otimes\Lambda_{1\mid Z_{x+1}}^{\mp,x+1} & \rightarrow\frac{2}{q}\times\frac{1}{2}\sum_{\sigma=\pm}\frac{1}{2}\left({\Lambda_{1\mid Z_{x}Z_{x+1}}^{\sigma}+\Lambda_{Z_{x,x+1}\mid Z_{x,x+1}}^{\sigma}}\right)+O(q^{-3}).
\end{align*}
The first mapping above shows that the cut can create red and blue particles from nothing in pairs, and is indeed equivalent to Eq.~\ref{eq:entanglementeom}. The second and third equation shows that the particles created can correlate with the direction of motion of the cut. We have further related equations showing that the cut can scatter a red to a blue particle and vice versa:
\begin{align*}
\Lambda_{Z_{x}\mid1}^{\pm,x}\otimes\Lambda_{1\mid1}^{\mp,x+1},\Lambda_{1\mid1}^{\mp,x}\otimes\Lambda_{Z_{x+1}\mid1}^{\pm,x+1} & \rightarrow\frac{2}{q}\frac{1}{4}\left({\Lambda_{Z_{x,x+1}\mid I}^{\pm}}+{\Lambda_{Z_{x}Z_{x+1}\mid Z_{x,x+1}}^{\pm}}+{\Lambda_{Z_{x,x+1}\mid I}^{\mp}+\Lambda_{I\mid Z_{x,x+1}}^{\mp}}\right)+O(q^{-3})\\
\Lambda_{1\mid Z_{x}}^{\pm,x}\otimes\Lambda_{1\mid1}^{\mp,x+1},\Lambda_{1\mid1}^{\mp,x}\otimes\Lambda_{1\mid Z_{x+1}}^{\pm,x+1} & \rightarrow\frac{2}{q}\frac{1}{4}\left({\Lambda_{I\mid Z_{x,x+1}}^{\pm}}+{\Lambda_{Z_{x,x+1}\mid Z_{x}Z_{x+1}}^{\pm}}+{\Lambda_{I\mid Z_{x,x+1}}^{\mp}+\Lambda_{Z_{x,x+1}\mid I}^{\mp}}\right)+O(q^{-3})
\end{align*}
Note also that
\begin{align*}
\Lambda_{Z_{x}\mid Z_{x}}^{\pm,x}\otimes\Lambda_{1\mid1}^{\mp,x+1},\Lambda_{1\mid1}^{\mp,x}\otimes\Lambda_{Z_{x+1}\mid Z_{x+1}}^{\pm,x+1}\rightarrow & \frac{2}{q}\frac{1}{4}\left({\Lambda_{Z_{x}Z_{x+1}\mid Z_{x}Z_{x+1}}^{\pm}}+{\Lambda_{Z_{x,x+1}\mid Z_{x,x+1}}^{\pm}}+{\Lambda_{1\mid1}^{\mp}}+{\Lambda_{Z_{x,x+1}\mid Z_{x,x+1}}^{\mp}}\right)+O(q^{-3})\\
\Lambda_{Z_{x}\mid1}^{\pm,x}\otimes\Lambda_{1\mid Z_{x+1}}^{\mp,x+1},\Lambda_{1\mid Z_{x}}^{\mp,x}\otimes\Lambda_{Z_{x+1}\mid1}^{\pm,x+1}\rightarrow & \frac{2}{q}\frac{1}{4}\left({\Lambda_{Z_{x}Z_{x+1}\mid1}^{\pm}+\Lambda_{Z_{x,x+1}\mid Z_{x,x+1}}^{\pm}+\Lambda_{1\mid Z_{x}Z_{x+1}}^{\mp}+\Lambda_{Z_{x,x+1}\mid Z_{x,x+1}}^{\mp}}\right)+O(q^{-3})
\end{align*}
the first of which shows that the cut can destroy particles in pairs as well. Finally
\begin{align*}
\Lambda_{Z_{x}\mid Z_{x}}^{\pm,x}\otimes\Lambda_{Z_{x+1}\mid Z_{x+1}}^{\mp,x+1} & \rightarrow\frac{2}{q}\times\frac{1}{2}\sum_{\sigma=\pm}\frac{1}{2}\left({\Lambda_{Z_{x}Z_{x+1}\mid Z_{x}Z_{x+1}}^{\sigma}}+{{\Lambda_{Z_{x,x+1}\mid Z_{x,x+1}}^{\sigma}}}\right)+O(q^{-3})\\
\Lambda_{Z_{x}\mid Z_{x}}^{\pm,x}\otimes\Lambda_{Z_{x+1}\mid1}^{\mp,x+1},\Lambda_{Z_{x}\mid1}^{\mp,x}\otimes\Lambda_{Z_{x+1}\mid Z_{x+1}}^{\pm,x+1} & \rightarrow\frac{2}{q}\frac{1}{4}\left({\Lambda_{Z_{x}Z_{x+1}\mid Z_{x,x+1}}^{\pm}}+{\Lambda_{Z_{x,x+1}\mid Z_{x}Z_{x+1}}^{\pm}}+{\Lambda_{Z_{x}Z_{x+1}\mid Z_{x,x+1}}^{\mp}+{\Lambda_{Z_{x,x+1}\mid I}^{\mp}}}\right)+O(q^{-3})\\
\Lambda_{Z_{x}\mid Z_{x}}^{\pm,x}\otimes\Lambda_{1\mid Z_{x+1}}^{\mp,x+1},\Lambda_{1\mid Z_{x}}^{\mp,x}\otimes\Lambda_{Z_{x+1}\mid Z_{x+1}}^{\pm,x+1} &\rightarrow\frac{2}{q}\frac{1}{4}\left({\Lambda_{Z_{x,x+1}\mid Z_{x}Z_{x+1}}^{\pm}}+{\Lambda_{Z_{x}Z_{x+1}\mid Z_{x,x+1}}^{\pm}}+{\Lambda_{Z_{x,x+1}\mid Z_{x}Z_{x+1}}^{\mp}+{\Lambda_{I\mid Z_{x,x+1}}^{\mp}}}\right)+O(q^{-3})
\end{align*}
Note that each of the last 16 terms involves the action of the circuit in the vicinity of the cut, and are correspondingly suppressed by a factor of $2/q$. Otherwise, these equations exhibit a rich set of behaviors. For example, the cut can create and absorb particles in pairs, and can convert a red particle to a blue particle and vice versa. The cut itself always appears to have an equal chance of moving to the left or right although its motion can correlate with changes to the red/blue populations. 

The full set of two site calculations above show that, up to $O(q^{-2})$ corrections, the Haar averaged dynamics are closed in the space of operators spanned by $\mathcal{F}_{\tau,\upsilon}(x)$ (see Eq.~\ref{eq:manifoldofsuperops}) where $\tau,\upsilon$ denote the possible configurations of red and blue particles alluded to in the text. Moreover, saving for the overall factor of $2/q$ at each global time step, the induced dynamics on this restricted space of operators is in fact Markovian leading to a stochastic process on the configuration space of cuts and blue and red particle configurations $\{x,\{\tau_y\},\{\upsilon_y\}\}$; i.e., ignoring the $2/q$ factor, the RHS coefficients in all the mappings listed below Eq.~\ref{eq:disting} add up to unity. This results in a probability distribution $p(x,\{\tau_y\},\{\upsilon_y\})$. As stated in the main text, one can take the mean field approximation
\begin{equation}
p(x,\{\tau_y\},\{\upsilon_y\})\propto e^{-\frac{x^{2}}{2l(t)^2}} e^{-\frac{1}{2l(t)^2}\sum_{y}\left(\tau_{y}+\upsilon_{y}\right)y^{2}},
\end{equation}
with $l(t)^2 = \kappa t$. Using this, and the definition of $\mathcal{F}_{\tau,\upsilon}(x)$, one find that the purity of an arbitrary initial state reads
\begin{equation}
\frac{\prod_{y}\left(1+\braket{Z_{y}}e^{-\frac{y^{2}}{2l(t)^2}}\right)^{2}+\prod_{y}\left(1-\braket{Z_{y}}e^{-\frac{y^{2}}{2l(t)^2}}\right)^{2}}{\prod_{y}\left(1+e^{-\frac{y^{2}}{2l(t)^2}}\right)^{2}+\prod_{y}\left(1-e^{-\frac{y^{2}}{2l(t)^2}}\right)^{2}}\label{eq:finalestimate}.
\end{equation}
For translation invarint product states, this expression simplifies to Eq.~\eqref{eq:large_q_purity_final} in the main text.

\section{Entanglement growth for various initial states}

Here we complement the results presented in the main text, where we averaged over different initial product states, with further data on various initial states. We start by considering the U$(1)$ symmetric random circuit at different values of $q$ and show that the annealed average entropy, $S_2^{(a)}$, behaves similarly for the different states defined around Eq.~\eqref{eq:entanglement_saturation} in the text.

\begin{figure}[h!]
	\centering
	\includegraphics[width=0.75\columnwidth]{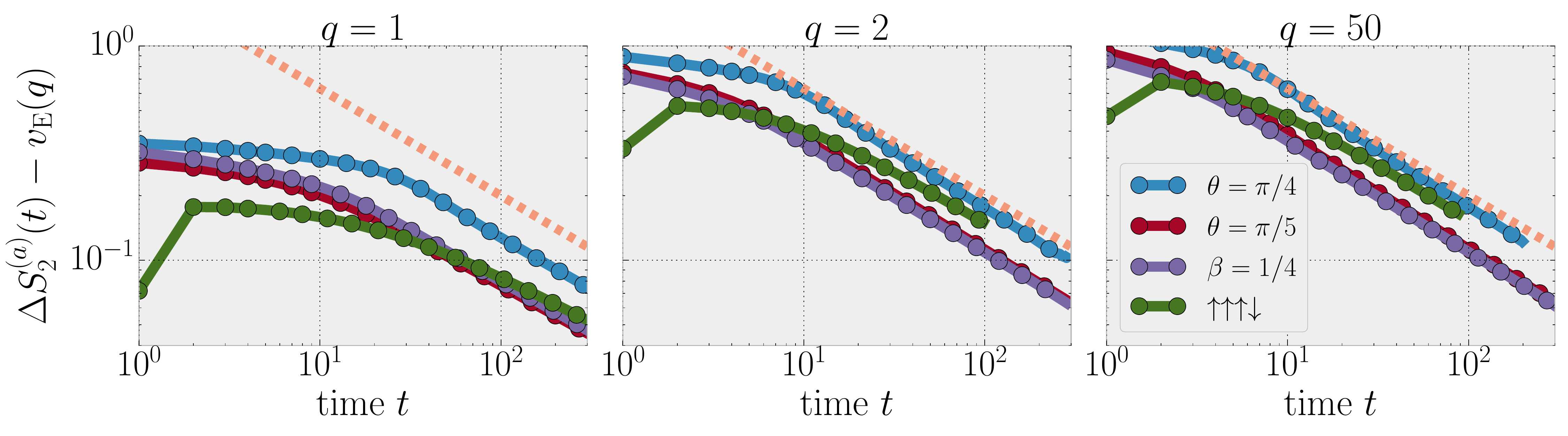}
	\caption{Time derivative of the annealed average R\'enyi entropy in the random circuit at different values of $q$, after subtracting the constant term, $v_\text{E}(q)$, associated to the non-conserved degrees of freedom. All states show a decay of the form $\propto t^{-1/2}$ at long times, indicating diffusive growth.} 
	\label{fig:RC_all}
\end{figure}

In the spin $1/2$ circuit, our results are consistent with the predicted $\propto\sqrt{t}$ growth of the R\'enyi entropy for all all initial states. The same is true at finite $q$ after subtracting the linear $v_\text{E}(q)t$ term coming from the additional degrees of freedom that do not carry any conserved charge. This is shown in Fig.~\ref{fig:RC_all}. We find that the times needed for N\'eel-like states to cross-over to sub-ballistic growth are longer then the time scales for other states that are not global eigenstates of the conserved quantity. This is in agreement with our heuristic argument, presented in the main text: the rare events leading to slow growth are already present for the latter states, while they have to be dynamically generated in the former.

Note that after subtracting $v_\text{E}t$ from the entanglement, there are only small differences between the curves for different $q$ (especially for $q\geq 2$). This further reinforces our interpretation that we can associate this quantity primarily to the conserved spins, and that we are justified in extrapolating the analytical results of the $q=\infty$ model back to $q=1$.

\begin{figure}[h!]
    \centering
	\includegraphics[width=0.385\columnwidth]{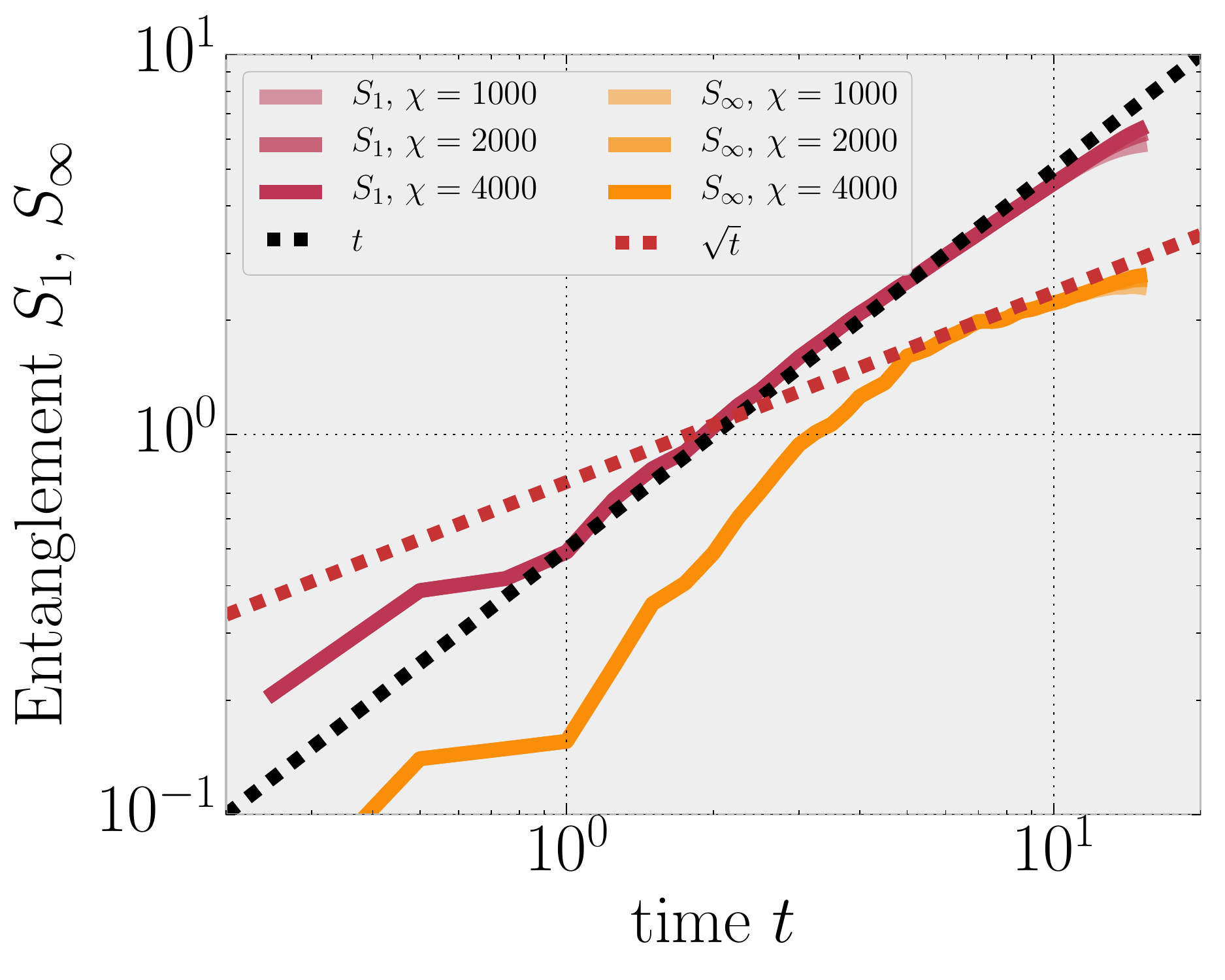}
	\includegraphics[width=0.385\columnwidth]{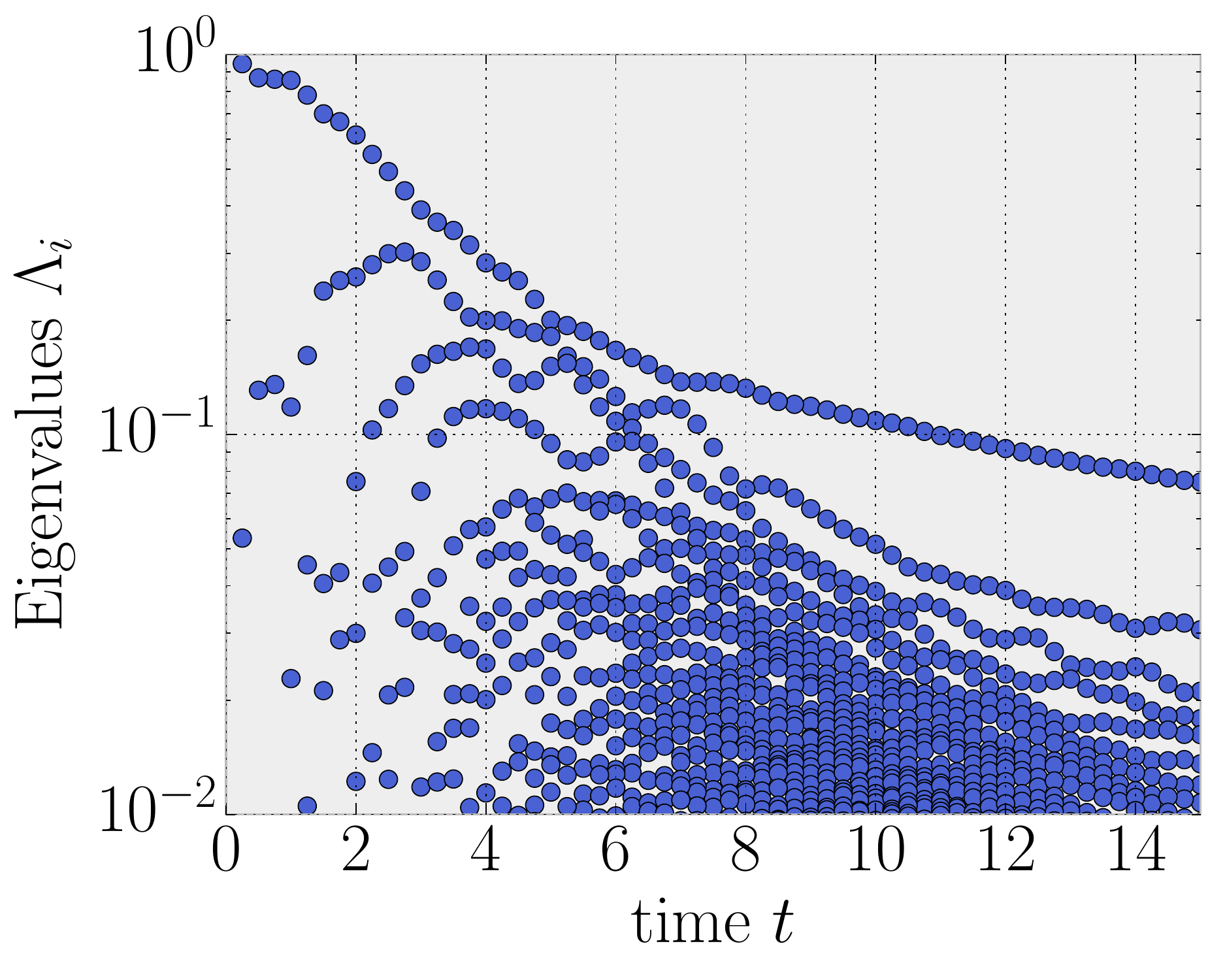}
	\caption{Entanglement growth in the tilted field Ising model, starting from an initial state polarized in the positive $Y$ direction, evolved with the iTEBD algorithm at various bond dimensions $\chi$. We observe a cross-over to sub-ballistic growth in the min-entropy $S_\infty$, as opposed to $S_1$, which remains ballistic (left). This shows up in the spectrum of the half-chain reduced density matrix, where the largest eigenvalue is separated from the bulk of the spectrum at long times (right).} 
	\label{fig:KimHuseY}
\end{figure}

We also consider the tilted field Ising model for a particular initial product state. To have a state which is both translation invariant and corresponds to infinite temperature, we take a state which is an eigenstate of the $Y$ Pauli operator on each site with an eigenvalue $+1$. Since both the state and the dynamics are translation invariant, we can time evolve directly in the thermodynamic limit, utilizing the infinite time-evolving block decimation (iTEBD) algorithm~\cite{VidalTEBD}. The times obtainable are then limited by the maximal bond dimension $\chi$. We find a behavior analogous to the one observed for random product states in the main text: $S_1$ grows approximately linearly, while $S_\infty$ curves over to an approximately diffusive growth at times $t\approx 10$ (left panel of Fig.~\ref{fig:KimHuseY}). One way to think about this phenomenon is as a separation of scales in the spectrum of the reduced density matrix, whose largest eigenvalues decay as $\Lambda_i \sim e^{-\sqrt{t}}$, while the majority of the eigenvalues are exponentially small, $\Lambda_i \sim e^{-t}$ (right panel of Fig.~\ref{fig:KimHuseY}).

\section{Hydrodynamic tails in the spin $1/2$ random circuit}\label{app:charge_fluctuations}

In this section we describe the behavior of various correlation functions in the U$(1)$ conserving random circuit model, lifting intuition from random circuit calculations performed previously in Refs. \cite{OTOCDiff1,OTOCDiff2}. Our results are in line with the standard lore of hydrodynamics: the space of local observables splits into an orthogonal sum of hydrodynamic and non-hydrodynamic variables. In a short range correlated initial state, the former can decay according to a power law at long times and have diffusive spatiotemporal behavior, while the latter always
decay exponentially quickly. An advantage of our approach, compared to the usual scaling arguments~\cite{Rosch13}, is that we are able to make statements about the differences in behavior for different homogenous initial states; something we expect to be of independent interest.

We start with an exact rewriting of the circuit evolved
$Z_{r}$ operator
\begin{equation}\label{eq:ansatz}
Z_{r}(t)=D_{r}(t)+B_{r}(t),
\end{equation}
where $D_{r}(t)=\sum_{r'}K_{rr'}\left(t\right)Z_{r'}$ and $K_{rr'}(t)$
is a lattice diffusion kernel, and where
\begin{equation}\label{eq:Bdetails}
B_{r}(t)=\frac{1}{2}\sum_{\tau=1}^{t}\sum_{x=\tau\mod2}\left(\partial_{x}K_{rx}(\tau)\right)\Gamma_{x}(t;\tau),
\end{equation}
with $\Gamma_{x}(t;\tau)\equiv\left(Z_{x}-Z_{x+1}\right)(t;\tau)$.
No approximations have been made thus far. The operator weight in
$D_{r}$ is $\sum_{r'}K_{rr'}^{2}=O(t^{-1/2})$, while that in $B_{r}$
is $1-O(t^{-1/2})$~\cite{OTOCDiff1,OTOCDiff2}. We will assume that $\Gamma_{x}(t;\tau)$ grows ballistically, so that it is effectively a random superposition of
operators of typical radius $v_{B}\left(t-\tau\right)$. This assumption
can only be an approximation to the full story, in part because $Z_{x}-Z_{x+1}$
is itself a hydrodynamical variable. However, it turns out that
the expectation values of this tail is either exactly zero ($\theta$-states) or decays exponentially in time (N\'eel-like states), so we are justified
in ignoring it.

In considering the growth of fluctuations, we also consider operators
of the form $Z_{r}(t)Z_{s}(t)$. These we may similarly write exactly
as
\begin{equation}\label{eq:twopt}
Z_{r}(t)Z_{s}(t)=\sum_{r's'}K_{rs}^{r's'}(t)Z_{r'}Z_{s'}+B_{rs}(t).
\end{equation}
The random circuit calculation demonstrates that, under averaging,
the two operators engage in single file diffusion, which we take to
have kernel $K_{rs}^{r's'}(t)$. $B_{rs}(t)$ is again assumed to be a ballistically
spreading and effectively random superposition of operators. In the
following we explain how to estimate the various correlation functions
presented in Table~\ref{tab:hydro_tails}, using operator spreading intuition. 

\paragraph{One point functions}

The average behavior of $\left\langle Z_{r}(t)\right\rangle $ is
simple to compute for various states through substitution of Eq. \ref{eq:ansatz}.
This leads to two expressions $\left\langle D_{r}(t)\right\rangle $
and $\left\langle B_{r}(t)\right\rangle $. The former is independent
of time for translation invariant states, and determined by the filling,
while for the Neel state it decays exponentially with $t$ because
$\sum_{r'}(-1)^{r'}K_{rr'}\left(t\right)\sim e^{-\frac{\pi^{2}t}{2}}$.
The $B_{r}$ term is more involved, but on average expected to be
zero because $\Gamma_{x}(t;\tau)$ is a superposition of operators
with random signs. 

\paragraph{Fluctuations in one point functions}

The average behavior of $\overline{\left\langle Z_{r}(t)\right\rangle \left\langle Z_{s}(t)\right\rangle }$
can also be computed using \ref{eq:ansatz}. Using the results in
the previous paragraph, two terms survive the noise averaging, $\left\langle D_{r}(t)\right\rangle \left\langle D_{s}(t)\right\rangle $
and $\overline{\left\langle B_{r}(t)\right\rangle \left\langle B_{s}(t)\right\rangle }$
. The former term can be calculated using the paragraph above, while
the latter require a more involved discussion. Using

\begin{equation}
\overline{\left\langle B_{r}(t)\right\rangle \left\langle B_{s}(t)\right\rangle }=\frac{1}{2}\sum_{,\tau'\tau=1}^{t}\sum_{x,y=\tau\mod2}\partial_{x}K_{rx}(\tau)\partial_{y}K_{sy}(\tau)\overline{\left\langle \Gamma_{x}(t;\tau)\right\rangle \left\langle \Gamma_{y}(t;\tau)\right\rangle }\label{eq:disconn}
\end{equation}

To approximate this sum, we first note that, as $\Gamma_{y}(t;\tau)$
tends to grow ballistically, its expectation value on a short range
correlated state will tend to decay exponentially quickly (a string
of operators of length $R$ has a typical expectation value $e^{-\alpha R}$
on a product state). So the sum is dominated by $\tau,\tau'$ near
$t$. Moreover, due to the spatial randomness of the circuit, the
sign of the expectation values $\left\langle \Gamma_{x/y}(t;\tau)\right\rangle $
are uncorrelated unless $x=y$. Hence we approximate Eq.~\ref{eq:disconn}
by restricting the sum to $\tau',\tau=t-1,x=y$. At large times $t\gg\left|r-s\right|^{2}/D$,
the distinction between $r,s$ becomes unimportant, and the contribution
goes as $\int dx\left(\partial_{x}K_{rx}\right)^{2}\sim t^{-3/2}$.

\paragraph{Two point functions}

In order to compute $\overline{\left\langle Z_{r}(t)Z_{s}(t)\right\rangle }$
we substitute in Eq.~\ref{eq:twopt} and assume the ballistic terms
source random signs that cancel on averaging. This leads to precisely
one contribution $\sum_{r's'}K_{rs}^{r's'}(t)\left\langle Z_{r'}Z_{s'}\right\rangle $
where $K_{rs}^{r's'}(t)$ is the single file lattice diffusion propagator.
This is independent of time for translation invariant states, but
remarkably is found to decay as $\sim\frac{1}{\sqrt{t}}$ for the
Neel state. In the case of $\epsilon$ states we substitute $\left\langle Z_{r'}Z_{s'}\right\rangle \sim e^{-\left|r'-s'\right|/\xi}$;
for finite $\epsilon$, this will have the effect of restricting $\left|r'-s'\right|\leq\xi$
in the sum above. The leading behavior is obtained by setting $r'=s'$,
which gives $\sum_{r'}K_{rs}^{r'r'}(t)\sim t^{-1/2}$.

\begin{table}
\centering
\begin{tabular}{|c|c|c|c|c|}
\hline 
State & $\overline{\braket{\delta Z_{r}}}$ & $\overline{\braket{\delta Z_{r}} \braket{\delta Z_{s}}}$ & $\overline{\delta(\braket{Z_{r}Z_{s}})}$ & $|\delta S_2|$\tabularnewline
\hline 
\hline 
$\uparrow\downarrow$ & $O(e^{-\pi^{2}t/2})$ & $O(t^{-3/2})$ & $O(t^{-1/2})$ & $O(t^{-1})$\tabularnewline
\hline 
$\uparrow\uparrow\downarrow$ & $O(e^{-\pi^{2}t/3})$ & $O(t^{-3/2})$ & $O(t^{-1/2})$ & $O(t^{-1/2})$\tabularnewline
\hline 
$\theta$-state & $0$ & $O(t^{-3/2})$ & $0$ & $O(t^{-3/2})$\tabularnewline
\hline 
$\beta$-state & $0$ & $O(t^{-3/2})$ & $O(t^{-1/2})$ & $O(t^{-1})$\tabularnewline
\hline 
\end{tabular}
\caption{Summary of the saturation behavior of various circuit averaged correlation functions, their moments and the R\'enyi entropy at large times $t\gg\left|r-s\right|^{2}$ in the spin $1/2$ random circuit.}
\label{tab:hydro_tails}
\end{table}

\end{document}

%% file: main.bbl
%

%% file: main.bbl
\begin{thebibliography}{50}
\bibitem [{Har()}]{HardCoreNote}%
  \BibitemOpen
  \href@noop {} {}\bibinfo {note} {We confirmed numerically the same scaling
  for classical particles with hard core interactions, initiated in a N\'eel
  state.}\BibitemShut {Stop}%
\bibitem [{Fru()}]{FrustFreeNote}%
  \BibitemOpen
  \href@noop {} {}\bibinfo {note} {The similarity is most apparent for local
  frustration free Hamiltonians, where one can consider states that are ground
  states of all local terms in a region.}\BibitemShut {Stop}%
\bibitem [{Mar()}]{MarkNote}%
  \BibitemOpen
  \href@noop {} {}\bibinfo {note} {This difference between entropies is
  consistent with holographic results, see Ref.
  \onlinecite{Mezei18}}\BibitemShut {NoStop}%
\end{thebibliography}
